\documentclass[aps,prl,preprint,groupedaddress]{revtex4}
\usepackage{epsfig}
\usepackage{subfigure}
\usepackage{amssymb}
\usepackage{graphicx}
\usepackage{amsmath}
\usepackage{natbib}

\begin{document}

\title{Optical non-reciprocity in magnetic structures \\ related to high-$T_c$ superconductors}


\author{J. Orenstein}
\affiliation{Department of Physics, University of California,
Berkeley, California 94720, USA.} \affiliation{Materials Science
Division, Lawrence Berkeley National Laboratory, Berkeley,
California 94720, USA.}

\date{\today}


\begin{abstract}
Recent neutron scattering \cite{HTSC:BreakTimeRev_neutron_YBCO,HTSC:BreakTimeRev_neutron_Hg1201}, and optical measurements
\cite{HTSC:KerrEffect_YBCO, PolarKerr2} have detected evidence in underdoped cuprate
superconductors for a phase transition near the pseudogap onset
temperature $T^*$ to a time reversal-breaking state. The
neutron scattering indicates antiferromagnetic ordering, while it is often assumed that optical polarization rotation requires
at least a weak ferromagnetic component. In this note we identify several antiferromagnetic structures, compatible with neutron scattering data,
that allow intrinsic polarization rotation through the magnetoelectic effect.
\end{abstract}

\maketitle
High-$T_c$ superconductivity in the cuprates remains a controversial field after more than 20 years of intense research \cite{Bonn}.
One of the few ideas that has achieved broad consensus is that solving the puzzle of high-$T_c$ requires
first understanding the normal phase from which it evolves. Early in the study of these materials,
various probes, particularly thermodynamic ones, failed to detect a phase transition above the critical temperature
for superconductivity. Instead, the most prominent feature of the normal state is the
loss with decreasing temperature of states at the Fermi energy - the pseudogap phenomenon \cite{pseudoreview}. The lack of
evidence for a phase transition suggested that the opening of the pseudogap
reflects a crossover, rather than the appearance of a new phase
with a distinct symmetry.  Subsequently, careful experiments revealed that in certain cuprates, at doping levels close
to 1/8, there were indeed phase transitions to states with true charge and spin density order \cite{stripe:neutron:TranquadaStripe1st}.
Thus the possibility exists that proximity to these translational symmetry breaking phases controls the physics in cuprates
that do not manifest this form of static order.

More recently, striking results have been obtained from neutron scattering experiments
\cite{HTSC:BreakTimeRev_neutron_YBCO,HTSC:BreakTimeRev_neutron_Hg1201}
 on a variety of underdoped cuprates that point to a well-defined
phase transition to a state with broken time reversal symmetry. The phase transition occurs at
temperatures associated with the appearance of the pseudogap state as determined by probes such as ARPES, NMR, and optical conductivity \cite{pseudoreview}.
Magnetic neutron scattering appears at wavevectors that coincide with reciprocal lattice vectors,
indicating that time-reversal breaking occurs without
loss of translational symmetry.  Furthermore, no scattering is observed when the in-plane component of the scattering vector is zero, suggesting that
the net moment of the unit cell is zero. Another significant finding is that the moments are not perpendicular to the Cu-O plane, but are are canted
at an angle of about 45 degrees. From these observations it can be concluded that antiferromagnetically-aligned magnetic moments exist
within each unit cell, whose origin could be the orbital currents
proposed by Varma \cite{HTSC:CompetingOrders_VarmaLoops}, electron spins,
or some combination of the two.

In another set of elegant experiments \cite{HTSC:KerrEffect_YBCO, PolarKerr2} performed on cuprate superconductors,
the Kapitulnik group has detected a very small
difference in the reflection coefficient of left and right circularly polarized light (Kerr effect)
using an ultrasensitive Sagnac interferometer.
Kerr rotation is a direct manifestation of broken time-reversal, and therefore its appearance tends to support the neutron scattering findings.
Although the experiments have yet to be performed on the same sample, it seems that the onset temperature of the Kerr rotation,
$T_K$, is systematically lower than the onset of antiferromagnetic ordering, $T_{AF}$, as observed by neutron scattering. One possible explanation is
that the Kerr rotation results from "weak ferromagnetism," a net moment induced in a pair of oppositely oriented
spins by a small antisymmetric canting of the spin direction. It is possible that weak ferromagnetism sets in at a temperature below $T_{AF}$,
triggered by the onset of some form of charge ordering, such as formation of a nematic state \cite{nematictheory,STM1,STM2,neutron,Nernst,STM3}.

The purpose of this note is to discuss a different, intrinsic, mechanism that generates Kerr rotation without requiring weak ferromagnetism.
This mechanism requires that time-reversal symmetry (T) and inversion symmetry (I) are broken, although the product T*I may be a symmetry operation.
This is in fact the symmetry of a pattern of intra-unit cell loop currents described by Simon and Varma \cite{ARPESdichroism}.
On the other hand, if the moments are associated
with spins that reside on the O atoms, inversion symmetry is not broken and this mechanism does not operate.  As we will see below, the
proposed mechanism also fails to operate if the moments are aligned parallel to the normal vector of the Cu-O plane, as predicted in the basic
form of the loop-current theory.  What is required is that the moments be tipped from the normal, as is in fact observed experimentally.

As pointed out previously \cite{symmetryconsideration}, a state in which T and I broken are broken by intra-unit cell currents can be in the
class of materials that exhibit magnetoelectric phenomena.
In these materals generation of magnetization by electric fields, and electrical polarization by magnetic fields, is allowed by symmetry.
In a classic experiment, Krichevtsov et al. \cite{Krichevtsov} observed Kerr rotation in the model magnetoelectric antiferromagnet Cr$_2$O$_3$.
This rotation is referred to as non-reciprocal, meaning that it changes sign in domains in which all the spins are flipped, i.e.
the sense of time is reversed.  The existence of non-reciprocal optical rotation in magnetoelectrics such as Cr$_2$O$_3$ had been predicted
theoretically many years ago \cite{Brown,Hornreich}, although the magnitude was thought to be extremely small.  Experiments showed that the rotation is
reasonably large and readily observable, of order $10^{-4}$, illustrating the well-known fact that symmetry arguments are quite powerful
in predicting the existence of exotic effects, but not necessarily their magnitude.  More recently, microscopic theories that include spin-orbit coupling
have brought theory and experiment into agreement \cite{cr2o3microscopic1,cr2o3microscopic2,cr2o3microscopic3}.

Motivated by these experiments, there has been considerable progress in developing a macroscopic theory of reflection
from magnetoelectrics \cite{GRtheory,Gridnev}.
As a result of this work, there exists a theoretical approach to calculate reflection coefficients from macroscopic property tensors
that is consistent with the requirements imposed by time-reversal and inversion symmetries \cite{Shelankov,Rojo}.  Below, we use
the theoretical approach described in Ref. \cite{GRtheory} to calculate the reflection matrix for several antiferromagnetic structures that appear
consistent with the neutron
scattering data as it currently exists.  Of these structures, it emerges that several give rise to a non-zero Kerr effect.  A phase transition
from a Kerr-inactive to Kerr-active antiferromagnetic state at $T_K$ is a scenario that could account for the onset of optical rotation at a
temperature below the onset of magnetic neutron scattering.

The macroscopic theory of magnetoelectric optics requires consideration of magnetization, $M$, induced by electric fields, $E$, which yields
effects at magnetic dipole order. The magnetoelectric tensor $t_{\alpha\beta}$ that relates $M$ to $E$ is a second-rank axial tensor that changes sign
under reverse of spin and current, that is a (c) tensor in the notation of Birss \cite{Birss}.  However, it turns out that for the theory to be consistent
with time-reversal, electric quadrupole effects must be included as well. A third-rank polar (c) tensor, $S_{\alpha\beta\gamma}$,
relates polarization, $P$, to the time and spatial derivative of $E$, yielding effects at electric quadrupole order.
Although both $S_{\alpha\beta\gamma}$ and $t_{\alpha\beta}$ contain nonzero elements in all magnetoelectrics, the optical properties of
a given material depend sensitively on its magnetic point group.

In the following, we discuss the optical reflectivity associated with five inversion breaking magnetic structures that are relevant to experiments.
The first, sketched in Fig. 1(a), is the structure described in Ref. \cite{ARPESdichroism}, which belongs to the magnetic point group $mmm^\prime$.  This structure is
characterized by three mirror planes, two of which contain the Cu-O unit cell diagonals and the normal vector.  The third lies in the Cu-O plane.
Of the three, the plane defined by the diagonal that passes through the moments and the normal direction is an $m^\prime$ plane, that is, the symmetry
operation is mirror followed by reversal of the direction of the orbital currents. As we will see, nonreciprocal rotation is not
observable in the $mmm^\prime$ group.  This fact, and the evidence from neutron scattering that
the moments are not perpendicular to the Cu-O plane, lead us to consider structures with different
symmetry. In Fig. 1(b) the magnetic moments are tipped away from the normal
direction while remaining within the $m^\prime$ plane.  In Fig. 1(c) the moments
are again tipped away from the normal, but in the direction perpendicular to the $m^\prime$ plane.
The lowering of symmetry removes two of the three mirror planes in both cases.  In structure shown in Fig. 1(b) the $m^\prime$ plane remains,
as does a 2-fold rotation about the normal to this plane.  Thus it is associated with the magnetic point group $2/m^\prime$.
In contrast, the structure shown in Fig. 1(c) is characterized by the $2^\prime$/m group, as the $m$ operation remains, together with a 2-fold
rotation followed by inversion about the $m$ plane normal.

\begin{figure}
\label{fig:fig1}
\includegraphics[width=5in]{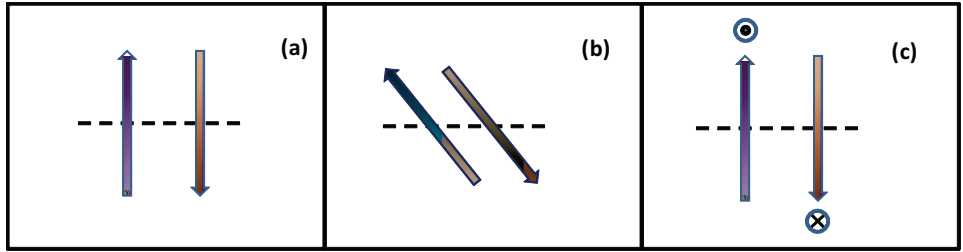}
\caption{Representation of possible symmetries of magnetic moments in a planar unit cell.  Dashed line represents the Cu-O plane, the plane of drawing
contains the normal and one of the diagonals of the CuO$_2$ cell. (a) moments perpendicular to the plane: point group $mmm^\prime$ (b) moments tipped
in the plane of the sketch: point group 2/$m^\prime$ (c) moments tipped out of the plane of the sketch: point group $2^\prime/m$.}
\end{figure}

\begin{figure}
\label{fig:fig1}
\includegraphics[width=3.5in]{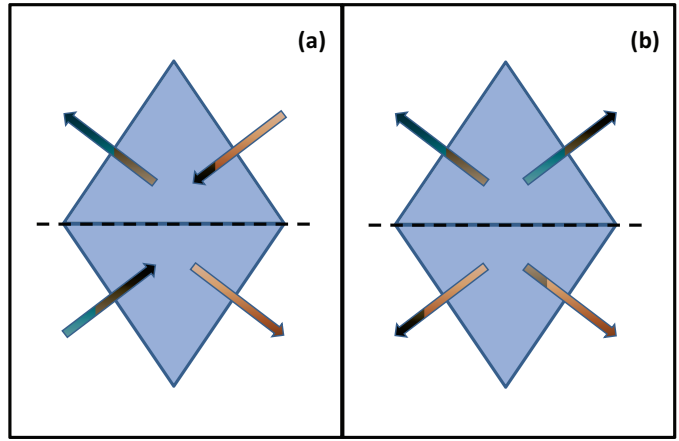}
\caption{Representation of possible symmetries of magnetic moments in a unit cell containing two apical oxygen atoms. (a) point group $mmm^\prime$
(b) point group $m^\prime m^\prime m^\prime$.}
\end{figure}

In another relevant structure studied by Weber \textit{et al.} \cite{giamarchi},
the orbital currents flow through the apical O, rather than the Cu atom at the center of the cell. This structure, represented by four
moments in Fig. 2(a), is also characterized by the point group $mmm^\prime$ \cite{symmetryconsideration}, and is therefore not a candidate
structure for nonreciprocal rotation.  However, reversing a pair of moments, as in Fig. 2(b), yields a point group $m^\prime m^\prime m^\prime$
structure, which will turn out to allow non-reciprocal optical rotation.  Although the moments in Fig. 2(b) lie in a plane, it may be possible that
the moments below the Cu-O plane could could be rotated by 90 degrees about $z$-axis with respect to the moments above.  Such a structure would
belong to the $2m^\prime m^\prime$ group, which has the same optical response as $m^\prime m^\prime m^\prime$.

The breaking of I and T allow for additional terms in the constitutive relations that the define of the optical response of
a magnetolectric medium.  In the formulation of Graham and Raab \cite{GRtheory}, these additional terms are,

\begin{equation}
M_\alpha=t_{\beta\alpha}E_\beta,
\end{equation}

\begin{equation}
P_\alpha=-\frac{1}{6}iS_{\alpha\beta\gamma}\nabla_\gamma E_\beta,
\end{equation}
where,
\begin{equation}
S_{\alpha\beta\gamma}\equiv a_{\alpha\beta\gamma}+a{\beta\gamma\alpha}+a_{\gamma\alpha\beta},
\end{equation} and,
\begin{equation}
t_{\alpha\beta}\equiv G_{\alpha\beta}-\frac{1}{3}\delta_{\alpha\beta}G_{\gamma\gamma}-\frac{1}{6}\omega\epsilon_{\beta\gamma\delta}
a_{\gamma\delta\alpha}.
\end{equation}
In the above $G_{\alpha\beta}$ is the second-rank axial (c) tensor and $a_{\alpha\beta\gamma}$ is the third-rank polar (c) tensor.
These additional terms, together with Maxwell's equations, lead to modifications of the boundary conditions and wave impedance. We consider
reflection at normal incidence, which we define as the $z$-direction. The boundary condition
on the tangential components of $H$ are,

\begin{equation}\label{bc}
\epsilon_{\alpha z\gamma}(H_{2\gamma}-H_{1\gamma})=\frac{1}{6}\omega S_{\alpha\gamma z}E_{1\gamma},
\end{equation}
while the condition of continuity of the tangential component of $E$ remains unchanged. The wave impedance, or coupling between $E$ and $H$ in the
medium, is described by the relation,

\begin{equation}\label{impedance}
H_\alpha=(\mu_0 c)^{-1}n\epsilon_{\alpha z\gamma}E_\gamma-t_{\beta\alpha}E_\beta.
\end{equation}

The coupling of $H_x$ to $E_x$ and $H_y$ to $E_y$ via a time-reversal odd tensor is the key ingredient that gives rise to non-reciprocal rotation,
as this is similar to the coupling in the presence of an external magnetic field parallel to the $z$ axis. Such coupling arises
from Eq. \ref{impedance} if the diagonal elements $t_{xx}$ and $t_{yy}$ are nonzero, or from Eq. \ref{bc} if the off diagonal components
$S_{xyz}$ and $S_{yxz}$ are nonzero.
To see which, if any, of the relevant magnetic structures generate a non-reciprocal rotation,
we evaluate tensor elements for the magnetic point groups under consideration,
\begin{equation}mmm': t_{xx}=t_{yy}=t_{xy}=0; S_{xyz}=S_{yxz}=0,
\end{equation}
\begin{equation}2'/m: t_{xx}=t_{yy}= 0; t_{xy}\neq 0; S_{xyz}=S_{yxz}=0,
\end{equation}
\begin{equation}m^\prime m^\prime m^\prime, 2m^\prime m^\prime: t_{xx}, t_{yy}\neq 0; t_{xy}=0; S_{xyz}, S_{yxz}\neq 0,
\end{equation}
\begin{equation}2/m^\prime: t_{xx}, t_{yy}\neq 0; t_{xy}=0; S_{xyz}, S_{yxz}\neq 0.
\end{equation}
From the above it is clear that non-reciprocal rotation can be observed in $2/m^\prime$, $m^\prime m^\prime m^\prime$,
$2m^\prime m^\prime$, but not in the other structures.

From the tensors $S_{\alpha\beta\gamma}$ and $t_{\alpha\beta}$, the reflectivity matrix, defined by $E_{r\alpha}=r_{\alpha\beta}E_{i\beta}$ (where
subscripts \textit{i} and \textit{r} denote incident and reflected respectively),
can be derived in the usual way from the boundary conditions and wave impedance relations.  The reflectivity matrix for the Kerr-active
structures is,

\begin{equation}\label{reflectionmatrix}
r=\large\begin{bmatrix}\frac{n-1}{n+1}&\frac{2(-\tau_{yy}+s_{xy})}{(n+1)^2}\\\frac{2(\tau_{xx}+s_{yx})}{(n+1)^2}&\frac{n-1}{n+1}
\end{bmatrix},
\end{equation}
where $\tau_{xx}\equiv \mu_0c t_{xx}$, $\tau_{yy}\equiv \mu_0c t_{yy}$, $s_{xy}\equiv(\omega \mu_0 c/6)S_{xyz}$, $s_{xy}\equiv(\omega \mu_0 c/6)S_{xyz}$.
In the above, we have made the approximation that the diagonal elements of $r_{\alpha\beta}$ are equal, as we consider a medium that would be tetragonal
if not for the magnetic structure.

The beauty of the Sagnac technique is that it is sensitive only to the time-reversal
breaking parameter, $\phi\equiv (r_{xy}-r_{yx})$.  For the point groups discussed above, $\phi=-2(\tau_{xx}+\tau_{yy}+s_{xy}-s_{yx})/(n+1)^2$.
From the permutation symmetry the third-rank polar tensor $a_{\alpha\beta\gamma}=a_{\alpha\gamma\beta}$ \cite{GRtheory} it follows that $s_{xy}-s_{yx}=0$.
Therefore $\phi\propto (\tau_{xx}+\tau_{yy})$, yielding the sensible result that the Sagnac signal is independent of the choice of crystallographic
axes. Finally, we note that the reflectivity matrix (Eq. \ref{reflectionmatrix}) is the same as obtained by Dzyaloshinskii \cite{anyon} in the context
of anyon theory, for the case of a bilayer cuprate in which the gauge fields of the two layers have opposite sign.  This state is
equivalent to a pair of antiferromagnetically aligned moments displaced along the $z$ axis, instead of the $x$ or $y$ axis as in Fig. 1(a).  That the
response is equivalent is consistent with symmetry considerations, as the corresponding magnetic point group is $m^\prime m^\prime m^\prime$, i.e. the
same as in Fig. 2(b).

Although the reflectivity matrix above will yield a Kerr effect, there are several issues that arise
when comparing Sagnac measurements with non-reciprocal rotation in magnetoelectrics.  One is the relatively small size of the effect, of order
0.1-1 microradians in cuprates \cite{HTSC:KerrEffect_YBCO} as compared with 100 microradians in Cr$_2$O$_3$ \cite{Krichevtsov}, and a second, perhaps related issue is "training," or alignment of domains.
As $\tau_{xx}+\tau_{yy}$ changes sign under reversal of spin direction, the optical rotation of a multidomain sample
averages to zero. The Sagnac
signal is found to increase when the sample is cooled in a magnetic field, an effect normally associated with the alignment of ferromagnetic, not
antiferromagnetic, domains.  However, although antiferromagnetic magnetoelectrics do not have free energy terms proportional to a static field in the
$z$-direction, $H_z^0$, there are terms proportional to $H_z^0 E_z^0$ in the Kerr-active point groups. If breaking of symmetry at the crystal
surfaces generates an $E_z^0$, the antiferromagnetic domains near the surface will be aligned by $H_z^0$.  If the rotation is confined to a
thin surface layer the relatively small size of the observed rotation could be explained.  Finally, a prediction that distinguishes magnetoelectric
from ferromagnetic training is that domains on the two opposite sides of the crystal are time-reversed in the magnetoelectric case.

To conclude, we have identified several magnetic structures that manifest non-reciprocal optical
rotation in the absence of weak ferromagnetism and are relevant to recent neutron scattering results.  This finding is
independent of whatever interactions
generate such structures and whether they arise from spin, orbital angular momentum, or both.  In carrying out the symmetry analysis
we have considered the highest symmetry (single-layer tetragonal) structure exemplified by the Hg 1201 compound.  Extending the analysis to
bilayer cuprates is more difficult, as it requires additional assumptions regarding the magnetic coupling between the two layers.  However,
if Kerr-active antiferromagnetic structures exist in the highest symmetry crystal structure, they will be found
in crystals with lower symmetry as well.

\begin{acknowledgments}
I would like to thank P. Bourges, J.C. Davis, J. Hinton, A. Kapitulnik, J. Koralek, S.A. Kivelson, A. Shekhter, and C.M. Varma for helpful discussions.
This work was supported by the Director, Office of Science, Office
of Basic Energy Sciences, Materials Sciences and Engineering
Division, of the U.S. Department of Energy under Contract No.
DE-AC02-05CH11231.
\end{acknowledgments}


\begin{thebibliography}{}

\bibitem{HTSC:BreakTimeRev_neutron_YBCO}
B. Fauqu$\acute{e}$ {\sl et al.}, \emph{Phys. Rev. Lett.} \textbf{96,} 197001 (2006).

\bibitem{HTSC:BreakTimeRev_neutron_Hg1201}
Y. Li {\sl et al.}, \emph{Nature} \textbf{455,} 372 (2008).

\bibitem{HTSC:KerrEffect_YBCO}
J. Xia {\sl et al.}, \emph{Phys. Rev. Lett.} \textbf{100,} 127002 (2008).

\bibitem{PolarKerr2}
A. Kapitulnik {\sl et al.}, \emph{New J. Phys.} \textbf{11} 055060 (2009).

\bibitem{Bonn}
D. A. Bonn, \emph{Nat. Phys.} \textbf{2,} 159 (2006).

\bibitem{pseudoreview}
T. Timusk and B. Statt, \emph{Rep. Prog. Phys.} \textbf{62,} 61 (1999).

\bibitem{stripe:neutron:TranquadaStripe1st}
J. M. Tranquada {\sl et al.}, \emph{Nature} \textbf{375,} 561 (1995).

\bibitem{HTSC:CompetingOrders_VarmaLoops}
C. M. Varma, \emph{Phys. Rev. B} \textbf{55,} 14554 (1997).

\bibitem{nematictheory}
S. A. Kivelson, E. Fradkin, V. J. Emery, \emph{Nature} \textbf{393,} 550 (1998).

\bibitem{STM1}
C. Howald {\sl et al.}, \emph{Proc. Natl. Acad. Sci.} \textbf{100,} 9705 (2003).

\bibitem{STM2}
Y. Kohsaka et al., {\sl et al.}, \emph{Science} \textbf{315,} 1380 (2007).

\bibitem{neutron}
V. Hinkov et al., {\sl et al.}, \emph{Science} \textbf{319,} 1380 (2008).

\bibitem{Nernst}
R.Daou et al., {\sl et al.}, \emph{Nature} \textbf{463,} 1380 (2010).

\bibitem{STM3}
M. J. Lawler et al., {\sl et al.}, \emph{Nature} \textbf{466,} 347 (2010).


\bibitem{ARPESdichroism}
M.E. Simon and C.M. Varma, \emph{Phys. Rev. Lett.} 89, 247003 (2002).

\bibitem{symmetryconsideration}
A. Shekhter and C. M. Varma, \emph{Phys. Rev. B} \textbf{80,} 214501 (2009).

\bibitem{Krichevtsov}
B.B. Krichevtsov {\sl et al.}, \emph{J. Phys.: Condens. Matter} \textbf{9,} 1863 (1993).

\bibitem{Brown}
W. F. Brown, Jr., S. Shtrikman, and D. Treves, \emph{J. Appl. Phys.} \textbf{34,} 1233 (1963).

\bibitem{Hornreich}
R. M. Hornreich and S. Shtrikman, \emph{Phys. Rev.} \textbf{171,} 1065 (1968).

\bibitem{cr2o3microscopic1}
V.N. Muthukumar, R. Valentí, and C. Gros, \emph{Phys. Rev. Lett.} \textbf{75,} 2766 (1995).

\bibitem{cr2o3microscopic2}
R. Valent\'{\i}, V.N. Muthukumar and C. Gros, \emph{Phys. Rev. B} \textbf{54,} 433 (1996).

\bibitem{cr2o3microscopic3}
M. Muto, Y. Tanabe, T. Iizuka-Sakano, and E. Hanamura, \emph{Phys. Rev. B} \textbf{57,} 9586 (1998).

\bibitem{GRtheory}
E.B. Graham and R.E. Raab, \emph{Phys. Rev. B} \textbf{59,} 7058 (1999).

\bibitem{Gridnev}
V. N. Gridnev, \emph{Phys. Rev. B}\textbf{51}, 13079 (1995).

\bibitem{Shelankov}
A.L. Shelankov and G. E. Pikus, \emph{Phys. Rev. B} \textbf{46,} 3326 (1992).

\bibitem{Rojo}
G. S. Canright and A.G. Rojo, \emph{Phys. Rev. B} \textbf{46,} 14078 (1992).

\bibitem{Birss}
R. R. Birss, \emph{Rep. Prog. Phys.} \textbf{26,} 307 (1963).

\bibitem{giamarchi}
C. Weber {\sl et al.}, \emph{Phys. Rev. Lett.} \textbf{102,} 017005 (2009).

\bibitem{anyon}
I. Dzyaloshinskii, \emph{Phys. Lett. A} \textbf{155,} 62 (1991).

\end{thebibliography}
\end{document}